\documentclass[12pt]{spieman}  % 12pt font required by SPIE;
\usepackage{amsmath,amsfonts,amssymb}
\usepackage{graphicx}
\usepackage{setspace}
\usepackage{tocloft}
\usepackage{hyperref}
\usepackage{color,soul}
\usepackage{lineno}

\title{A digital instrument simulator to optimize the development of hyperspectral systems: application for intraoperative functional brain mapping}

\author[a*]{Charly Caredda}
\author[b]{Frédéric Lange}
\author[c]{Luca Giannoni}
\author[d]{Ivan Ezhov}
\author[e]{Thiébaud Picart}
\author[e]{Jacques Guyotat}
\author[b]{Ilias Tachtsidis}
\author[a]{Bruno Montcel}

\affil[a]{Univ Lyon, INSA-Lyon, Université Claude Bernard Lyon 1, UJM-Saint Etienne, CNRS, Inserm, CREATIS UMR 5220, U1294, F69100, Lyon, France}
\affil[b]{Department of Medical Physics and Biomedical Engineering, University College London, UK}
\affil[c]{Department of Physics and Astronomy, University of Florence, Italy}
\affil[d]{Technical University of Munich, Germany}
\affil[e]{Service de Neurochirurgie D, Hospices	Civils de Lyon, Bron, France}

\cftpagenumbersoff{figure}
\cftpagenumbersoff{table} 
\begin{document} 
\maketitle

\begin{abstract}
	
	%\noindent \textbf{Significance:} Intraoperative optical imaging is a localization technique for the functional areas of the	human brain cortex during neurosurgical procedures. These areas are assessed by monitoring the oxygenated ($HbO_2$), deoxygenated hemoglobin ($Hb$) and the oxidative state of cytochrome-c-oxidase ($oxCCO$). A robust quantification of hemodyanmic and metabolism markers is complicated to perform during neurosurgery due to the critical context of the operating room. In actual devices, the inhomogeneities of the optical properties of exposed brain cortex are poorly taken into consideration, which introduce quantification errors of biomarkers of brain functionality. Moreover, the choice of the best spectral configuration is still based on an empirical approach.
	
	\noindent \textbf{Significance:} Intraoperative optical imaging is a localization technique for the functional areas of the	human brain cortex during neurosurgical procedures. These areas can be assessed by monitoring cerebral hemodynamics and metabolism. A robust quantification of these biomarkers is complicated to perform during neurosurgery due to the critical context of the operating room. In actual devices, the inhomogeneities of the optical properties of exposed brain cortex are poorly taken into consideration, which introduce quantification errors of biomarkers of brain functionality. Moreover, the choice of the best spectral configuration is still based on an empirical approach.
	
	\noindent \textbf{Aim:} We propose a digital instrument simulator to optimize the development of hyperspectral systems for intraoperative brain mapping studies. This simulator can provide a realistic modelling of the cerebral cortex and the identification of the optimal wavelengths to monitor cerebral hemodynamics (oxygenated $HbO_2$ and deoxygenated hemoglobin $Hb$) and metabolism (oxidized state of cytochromes b, c and cytochrome-c-oxidase $oxCytb$,  $oxCytc$ and $oxCCO$).
	
	%This simulator is based on a realistic digital phantom of an exposed cortex computed with Monte Carlo simulations. 

	\noindent \textbf{Approach:} The digital instrument simulator is computed with white Monte Carlo simulations of a volume created from a real image of exposed cortex. We developed an optimization procedure based on a genetic algorithm to identify the best wavelength combinations in the visible and near infrared range to quantify concentration changes in $HbO_2$, $Hb$, $oxCCO$ and the oxidized state of cytochrome b and c ($oxCytb$ and $oxCytc$).
	
	\noindent \textbf{Results:} The digital instrument allows the modelling of intensity maps collected by a camera sensor as well as images of pathlength to take into account the inhomogeneities of the optical properties. The optimization procedure helps to identify the best wavelength combination of $18$ wavelengths that reduce the quantification errors in $HbO_2$, $Hb$, $oxCCO$ of $61 \%$, $29\%$ and $82\%$ compared to the gold standard of 121 wavelengths between $780$ and $900$ nm. The optimization procedure does not help to resolve changes in cytochrome b and c in a significant way but help to better resolve $oxCCO$ changes.
	
	\noindent \textbf{Conclusions:} We proposed a digital instrument simulator to optimize the development of hyperspectral systems for intraoperative brain mapping studies. This digital instrument simulator and this optimization framework could be used to optimize the design of hyperspectral imaging devices.
	
\end{abstract}

% Include a list of up to six keywords after the abstract

% Include email contact information for corresponding author
{\noindent \footnotesize\textbf{*}Charly Caredda,  \linkable{charly.caredda@creatis.insa-lyon.fr} }

\begin{spacing}{2}   % use double spacing for rest of manuscript

\section{Introduction}

%General context
Non-invasive functional brain mapping is a technique that allows the locating of functional areas of the patient’s brain. This technique is used during brain tumor resection surgery to indicate to the neurosurgeon the cortical tissues which should not be removed to avoid motor, speech and cognitive impairments. Functional magnetic resonance imaging (fMRI) \cite{fMRI} is the preoperative gold standard for identifying the patient’s functional areas. However, after the patient’s craniotomy, a brain shift invalidates the relevance of neuronavigation to localize functional areas during surgery \cite{brain_shift}. To avoid localization errors, intraoperative MRI has been suggested \cite{intra_fMRI}, but it is costly and time consuming and, above-all, intraoperative MRI is available only in a very restricted number of neurosurgical centers. During neurosurgery, electrical brain stimulation (EBS) is the gold standard, but this technique is mainly limited by its low spatial resolution ($\approx5$ mm \cite{EBS_resolution}) and has the potential risk to trigger epileptic seizures. This technique allows a robust and reliable detection of brain functions, but could be traumatic for the patient, by inhibiting transiently certain brain functions such as speech \cite{langage_EBS}.\\

%Optical imaging and biomarkers
Optical imaging is a perfect complement to EBS since this technique is contact-less, non-invasive, non-ionizing and has a low traumatic impact for the patient. Indeed using optical imaging, the paradigms used to stimulate specific brain functions are similar to those used for fMRI \cite{caredda_neuroimage}. The analysis of the light absorbance allows to monitor the brain activity (motor or sensory tasks for example) with quantification of chromophores in the brain cortex: the concentration changes in oxy- ($\Delta C_{HbO_2}$) and deoxygenated hemoglobin ($\Delta C_{Hb}$) \cite{Chance,caredda,caredda2,caredda_resting_state,caredda_neuroimage} and cytochrome-c-oxidase ($\Delta C_{oxCCO}$) \cite{proc_caredda3,luca_oxCCO,review_oxCCO}, a mitochondrial marker of metabolism. $CCO$ is an enzyme in the mitochondria of the neuronal cells and is the terminal electron acceptor in the electron transport chain. Total $CCO$ concentration changes are slow and not correlated with brain activation. However, during cerebral activity, there are symmetrical variations in its oxidized and reduced states. Near infrared spectroscopy (NIRS) studies showed that it is possible to resolve changes in the oxidation state of $CCO$ using broadband spectroscopy procedures \cite{review_oxCCO,review_oxCCO2,CYRIL} using the difference spectrum between the oxidized and reduced species. $CCO$ is not the only enzyme that takes parts in the electron transport chain, other chromophores are involved such as cytochrome b and c ($Cytb$, $Cytc$). In \textit{in vivo} NIRS studies, $oxCCO$ changes are measured with near infrared light. To observe and measure other chromophores, visible light is required. However, this spectral range is difficult to use \textit{in vivo} due to the poor penetration of the light in the tissue. During neurosurgery, the cerebral cortex is exposed, making it possible to measure cerebral hemodynamics and metabolism at visible and near infrared wavelengths.\\

%Phantoms
A robust quantification of hemodynamic and metabolism markers is complicated to perform during neurosurgery due to the critical context of the operating room, which makes the calibration of optical devices more complex. To overcome this issue, tissue-simulating objects are required for the development of medical imaging devices. These so-called "phantoms" may be used to evaluate, optimize, compare or control imaging systems \cite{review_phantom}. Some phantom recipes based on intralipid and blood \cite{liquid_blood} and cytochrome contained yeast \cite{TR_system, Phantom_Fred} have emerged to reach that goal. The purpose of these devices is to evaluate the capacity of the acquisition system to follow the variations of oxygenation of the liquid. However, it is not possible to model hemodynamic responses similar to those that occur in the brain. Since, these phantoms are based on intralipid and blood, it is not possible to create heterogeneous phantoms that mimic the vascular network of the brain. For this reason, digital brain phantoms may be especially adapted to model hemodynamic and metabolic changes similar to those occurring during cerebral activity \cite{numeric_phantom_caredda,dynamic_model_Fantini} as well as 3D heterogeneous grey matter volumes \cite{3D_Heterogeneous_Phantom}.\\

%Partial volume effects and optimized wavelength
With wide-field imaging devices implying the use of a camera and an homogeneous illumination of the cerebral cortex, the biomarkers of brain hemodynamics and metabolism are quantified with the modified Beer-Lambert law \cite{caredda_neuroimage,luca_oxCCO}. This spectroscopic technique is however subject to quantification errors such as crosstalks \cite{review_oxCCO} and partial volume effect \cite{partial_vol_effect}. The crosstalk is defined as a concentration change of a chromophore that induce a change in another chromophore \cite{CYRIL}. The partial volume effect is wavelength dependent and refers to the impact of a focal change in the optical properties of a portion of the tissue on the surrounding tissue. To reduce the impact of the partial volume effect, the optical mean path length used in the modified Beer-Lambert law needs to be precisely estimated to consider the inhomogeneities of the optical properties \cite{caredda,3D_Heterogeneous_Phantom}. To reduce the quantification errors, a great number of wavelengths is recommended \cite{fNIRS_algo}. In \textit{in vivo} applications, a broadband spectra in the near infrared range between $780$ nm and $900$ nm is adapted for monitoring changes in $HbO_2$, $Hb$ and $oxCCO$ due to the differences between the molar extinction spectra of $HbO_2$, $Hb$ and the strong peak in the molar extinction spectra of $oxCCO$, as well as the deep penetration of the light into the tissue and low scattering effects. For acquiring broadband spectra for wide-field applications, hyperspectral imaging techniques need to be used \cite{review_HSI,review_HSI2}. These broadband systems require a complex instrumentation that implies a compromise in the spatial, spectral and temporal resolution of the acquisitions. Thus, the number of spectral bands in these systems is often limited by the experimental devices. Hyperspectral devices reconstruct spatially and spectrally resolved images. It generates a large amount of data that could be difficult to process in real time. For these reasons, specific wavelengths have been proposed in the literature to monitor hemodynamic an metabolic changes. Bouchard et al. \cite{Bouchard} used a monochrome camera combined with two wavelengths illumination ($470$, $530$ nm) to monitor hemodynamic changes in a mouse model during cerebral activity. In the same way, White et al. \cite{White} used two more wavelengths ($470$, $530$, $590$ and $625$ nm). Arifler et al. \cite{Arifler} proposed an optimization procedure to identify a reduced number of wavelengths to monitor $HbO_2$, $Hb$ and $oxCCO$ within the spectral range $780-900$ nm. Authors showed that a combination of $8$ wavelengths ($784$, $800$, $818$, $835$, $851$, $868$, $881$, and $894$ nm) allows a quantification errors less than $2 \%$ compared to a reference measurement (broadband spectra between $780$ and $900$ nm). \\

In this study, we developed a digital instrument simulator to optimize the development of hyperspectral systems for intraoperative brain mapping studies. This simulator can provide a realistic modelling of the cerebral cortex and the identification of the optimal wavelengths to monitor cerebral hemodynamics (oxygenated $HbO_2$ and deoxygenated hemoglobin $Hb$) and metabolism (oxidized state of cytochrome b, c and cytochrome-c-oxidase $oxCytb$,  $oxCytc$ and $oxCCO$).

\section{Material and methods}

\subsection{Digital instrument simulator} \label{sec:digital_instrument_simulator}

\begin{figure}[H]
	\centering
	\includegraphics[width=1.0\linewidth]{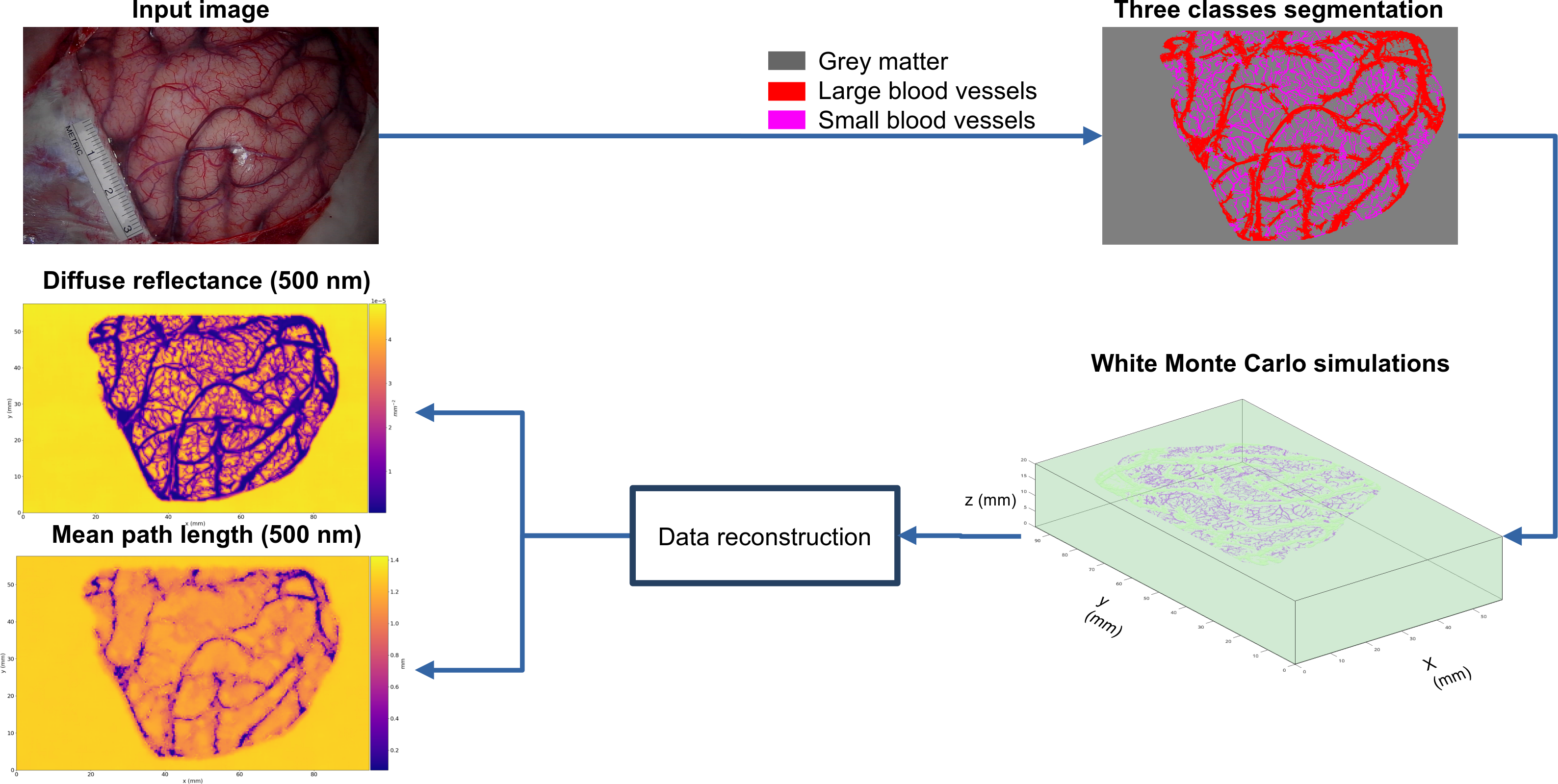}
	\caption{Flowchart of digital instrument simulator.}
	\label{fig:schema}
\end{figure}

The digital instrument simulator is based on a realistic digital phantom of an exposed cortex computed with Monte Carlo simulations. The code is publicly available on \href{https://github.com/CCaredda/White-Monte-Carlo/}{GitHub}. The Monte Carlo framework is composed in several steps, see Fig.~\ref{fig:schema}. First a color image of the exposed cortex was taken during the neurosurgery. This image was segmented into three classes (grey matter, large and small blood vessels). A brain volume was then modelled and a white Monte Carlo approach was used to estimate the partial path length, the position and the angle of exiting photons. Then we developed a C++ software to reconstruct temporal perturbations of the absorption coefficient using only one simulation.

\subsubsection{Data collection}

A color image of human brain exposed cortex was collected during a brain tumor resection operation at the neurological center of the Pierre Wertheimer hospital in Bron (France). A $8$ bits image was acquired with a surgical microscope Leica M530 OHX ($1280 \times 720$ pixels with a resolution of $73$ $\mu m$). The experiment was approved by the local ethics committee of Lyon University Hospital (France) and the participating patients signed written consent.

\subsubsection{Image segmentation and brain volume modelling}

\begin{figure}[H]
	\centering
	\includegraphics[width=0.7\linewidth]{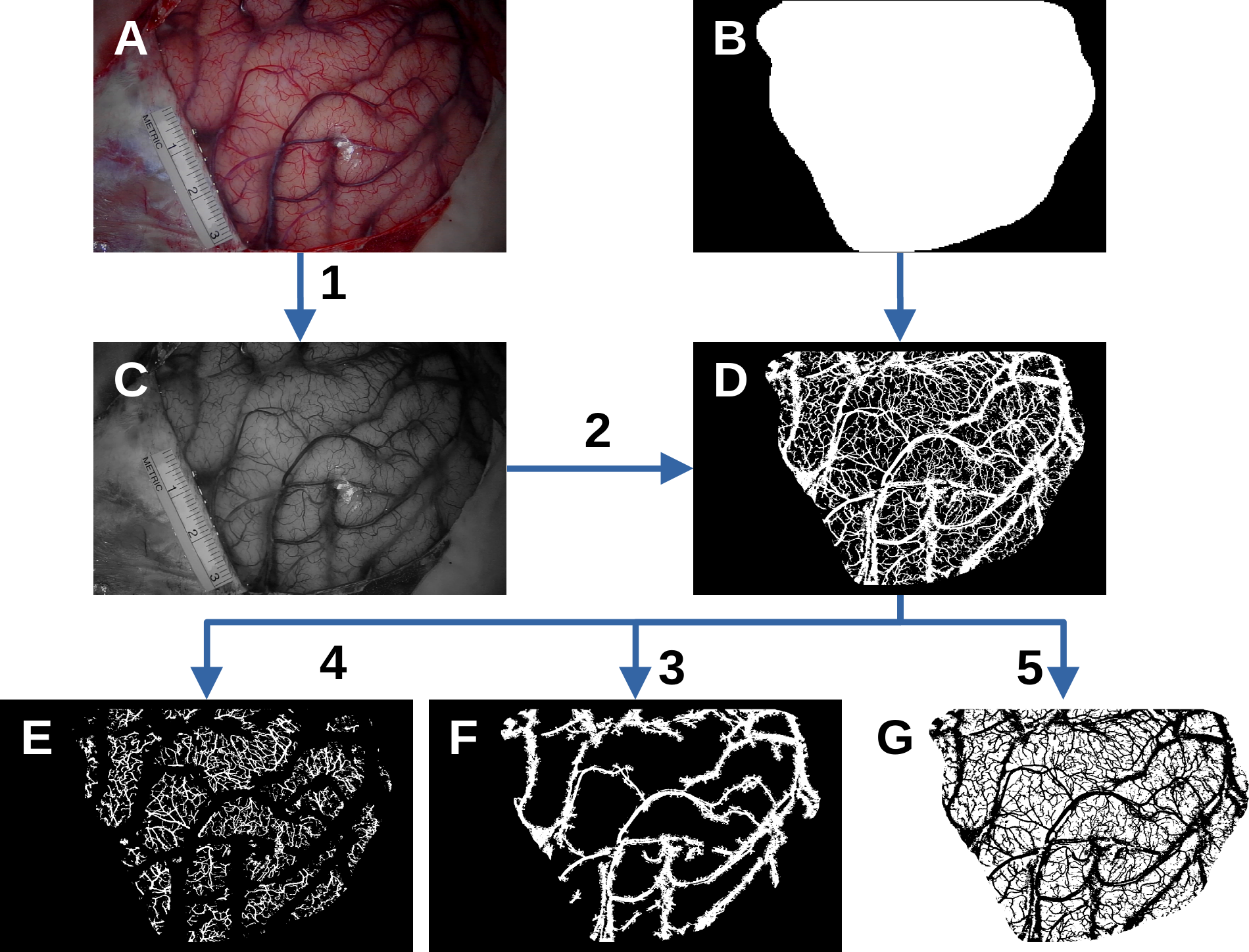}
	\caption{Segmentation of a color image (A) into masks of large blood vessel (F), small blood vessel (E) and grey matter (G). 1 - Conversion of color (A) into grayscale image (C). 2 - Gaussian adaptive thresholding to obtain the binary image (D). 3 - Computation of the large blood vessel mask (F). 4 - Computation of the small blood vessel mask (E). 5 - Computation of the grey matter mask (G).}
	\label{fig:segmentation}
\end{figure}

The color image was segmented into three classes (grey matter, large and small blood vessels) using morphological operations, see Fig.~\ref{fig:segmentation}.

\begin{enumerate}
	\item The color image (A) was converted into grayscale (C) using the Python library OpenCV (v4.8.1) \cite{OpenCV}.
	\item The grayscale image (C) was converted into a binary image (D) using a Gaussian adaptive threshold with a block size of $5 \times 5$ mm. The adaptive threshold was only applied on the exposed cortex by applying the mask (B). The latter is created manually. Saturated pixels due to specular reflection were removed from the image (D) with a simple thresholding of image (C). 
	\item The small vessels were removed from the binary image (D) using a morphological opening with a disk of $0.5$ mm diameter as a structuring element. Using the resulting image, a morphological closing was computed to get the image (F), the large blood vessel mask.
	\item A binary exclusive disjunction (xor) was computed between images (D) and (F) to get the small blood vessel mask (E).
	\item The mask of grey matter (G) corresponded to the rest of the pixels.
\end{enumerate}

%brain volume modeling
Once the image was segmented into three classes, we modelled the brain volume. 

\begin{figure}[H]
	\centering
	\includegraphics[width=1.0\linewidth]{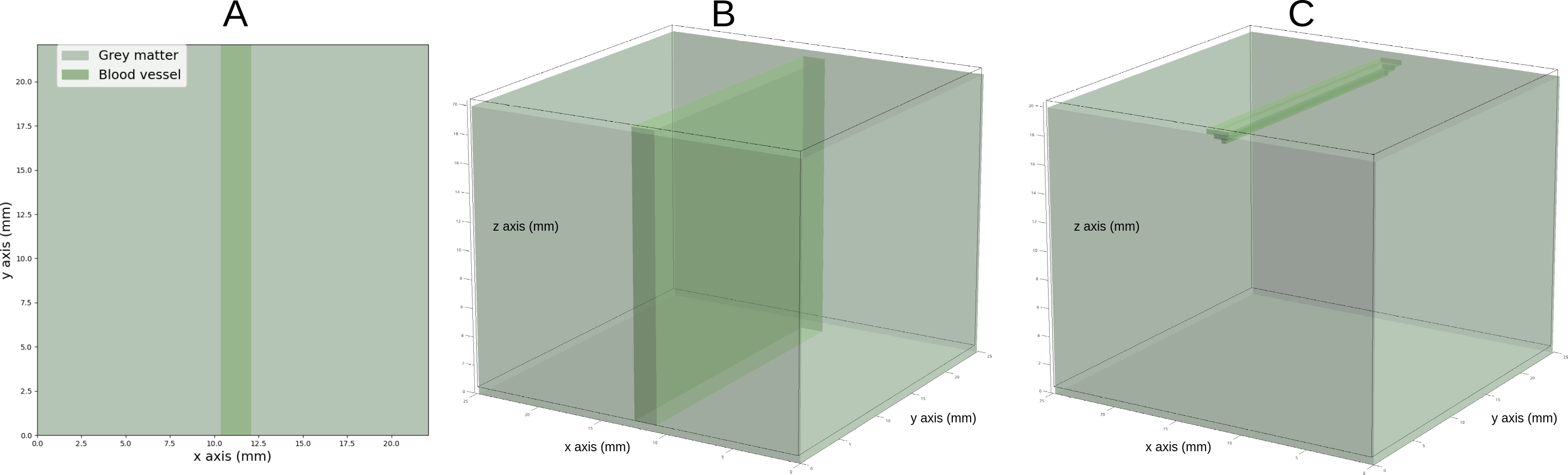}
	\caption{Steps for modelling a brain volume with blood vessels. A - Segmentation mask of the grey matter and the blood vessels. B- Replication of the binary segmentation masks along the vertical direction (z axis) on $2$ cm. C - Morphological erosion to create the blood vessel depth.}
	\label{fig:schema_bv_creation}
\end{figure}

First, the binary segmentation masks were encased in a larger image having the label of grey matter. This operation is performed to prevent a blood vessel label from appearing at the edges of the image. The objective is to avoid photons loss due to the boundary effects during the Monte Carlo simulations, see section \ref{sec:WMC}. The binary segmentation masks were expanded along the vertical direction (z axis) on $2$ cm to convert the binary images into binary volumes, see Fig. \ref{fig:schema_bv_creation} A and B. Then, we modelled the 3d blood vasculature with morphological erosion, see Fig. \ref{fig:schema_bv_creation} C. The depth of the blood vessels was calculated on the basis of the vessels diameter. The structuring element used for the erosion was set to $0$ pixels for $z=0$ (in pixels) and was increased of 1 pixel while increasing $z$ axis.

The binary volumes of the three classes were finally merged together with a final isotropic resolution of $73$ $\mu m$, see Fig.~\ref{fig:schema}.

\subsubsection{White Monte Carlo simulations} \label{sec:WMC}

Once the brain volume defined, we computed Monte Carlo simulations of light propagation in the domain using MCX software \cite{MCX}. A planar light source was placed on the top of the volume to homogeneously illuminate the brain volume. We set Fresnel reflection as boundary conditions for the top and bottom faces of the volume (i.e an exiting packet of photons is lost). Cyclic boundary conditions were applied for the sides of the volume to avoid photons loss due to the boundary effects. Each voxel of the volume included optical properties, see table \ref{tab:optical_prop}.

\begin{table}[H]
	\centering
	\begin{tabular}{c|c|c|c}
		& Grey matter & Large blood vessels & Small blood vessels \\
		\hline
		Absorption coefficient $\mu_a$ (mm$^{-1}$) & $0$ & $0$ & $0$ \\
		\hline
		Scattering coefficient $\mu_s$ (mm$^{-1}$) & $4.08 \left( \frac{\lambda}{500} \right)^{-3.089}$ \cite{optical_prop} & $2.2 \left( \frac{\lambda}{500} \right)^{-0.66}$ \cite{optical_prop} & $2.2 \left( \frac{\lambda}{500} \right)^{-0.66}$  \cite{optical_prop}\\
		\hline
		Anisotropy coefficient $g$  & $0.85$ \cite{g_brain} & $0.935$ \cite{prop_blood} & $0.935$ \cite{prop_blood} \\
		\hline
		Refractive index $n$  & $1.36$ \cite{n_brain} & $1.4$ \cite{prop_blood}  & $1.4$ \cite{prop_blood}\\
	\end{tabular}
	\caption{Optical properties used in the White Monte Carlo simulations. $\lambda$ denotes the wavelength in nm.}
	\label{tab:optical_prop}
\end{table}

For these simulations, we implemented a white Monte Carlo approach, the absorption coefficient was set to $0$ mm$^{-1}$. With this technique, the packets of photons were not affected by the absorption. For each wavelength (from $400$ nm to $1000$ nm by steps of $10$ nm), three outputs were stored for every packets of photons exiting the top face of the volume: the position of the exiting packet of photons $(x,y)$, its exiting angle and the partial pathlength (the length that each photon has spent in each classes of the domain).

\subsubsection{Image reconstruction}

%Reconstruction on the surface on the volume
Once the white Monte Carlo simulations are performed for each wavelength, diffuse reflectance and mean path length images can be reconstructed. Since a large amount of data is handled for each wavelength ($\approx 7$ Gb), a C++ software was developed for the reconstruction using CPU optimization. The advantage of the white Monte Carlo approach is that the absorption can be considered \textit{a posteriori}, by using the microscopic Beer-Lambert law \cite{micro_BLL}. Thus, when changing the absorption parameters of the volume (i.e., to simulate a brain activation for example), the simulation does not need to run again, speeding up the processing. In this case only one simulation per wavelength is required, considering that the scattering properties of the medium do not change. A scattering change can be considered but a new simulation with a new set of $\mu_s$ would need to be rerun. The absorption coefficient of the volume was computed with the chemical composition of the volume \cite{optical_prop}. These values were taken from the literature and corresponded to a nominal physiological condition \cite{cyt_values,caredda_neuroimage,opt_brain1,opt_brain2}, see table \ref{tab:mua}. A cerebral activation will be considered latter in the manuscript, see section \ref{sec:optim}. With our equipment, it is not possible to determine whether the blood vessels identified by the segmentation method are arteries or veins. For this reason, we considered two cases: (1) the large and small blood vessels are arteries, (2) the large and small blood vessels are veins.

\begin{table}[H]
	\centering
	\begin{tabular}{c|c|c|c}
		& Grey matter & Arteries & Veins  \\
		\hline
		$F_{Water}$ ($\%$) & $73$ & $55$ &$55$ \\
		\hline
		$F_{Fat}$ ($\%$) & $10$ & $1$ & $1$ \\
		\hline
		$C_{HbT}$ ($\mu M$) & $76$ & $2324$ & $2324$ \\ 
		\hline
		$SatO_2$ ($\%$) & $85$ & $98$ \cite{val_SatO2_vein_arteries} & $60$ \cite{val_SatO2_vein_arteries} \\
		\hline
		$C_{oxCCO}$ ($\mu M$) & $6.4$ & $0$ & $0$ \\
		\hline
		$C_{redCCO}$ ($\mu M$) & $1.6$ & $0$ & $0$ \\
		\hline
		$C_{oxCytb}$ ($\mu M$) & $2.37$ & $0$ & $0$ \\
		\hline
		$C_{redCytb}$ ($\mu M$) & $0.89$ & $0$ & $0$ \\
		\hline
		$C_{oxCytc}$ ($\mu M$) &$1.36$  & $0$ & $0$ \\
		\hline
		$C_{redCytc}$ ($\mu M$) & $0.68$ & $0$ & $0$ \\
	\end{tabular}
	\caption{Chemical composition of the modelled tissue}
	\label{tab:mua}
\end{table}

These images are reconstructed on a camera sensor that could be located on the tissue surface or outside the volume using a lens system. On the surface of the volume, the number of pixels is at most equal to that of the modelled surface. A spatial binning can be performed to increase the signal to noise ratio at the expense of spatial resolution.

To reconstruct the image on modelled camera sensor using a lens system, we used the transfer-matrix method \cite{transfer_matrix}. With this approach, we can consider key parameters of the optical system (focal length, working distance, size of the optics, size of the camera sensor). A lens with a focal length of $f_0$ (in mm) was placed at a distance $d0$ (in mm) from the surface of the tissue (along the axis $z$), and the sensor was located at a distance $s$ (in mm) from the lens (along the axis $z$). The optical system is modelled with a system matrix $\mathbb{S}$:

\begin{equation}
	\mathbb{S}  = \mathbb{T}_s \mathbb{L}_{f_0} \mathbb{T}_{d0} =
	\begin{pmatrix}
		1 & s \\
		0 & 1 
	\end{pmatrix}
	\begin{pmatrix}
		1 & -\frac{1}{f_0} \\
		0 & 1 
	\end{pmatrix}
	\begin{pmatrix}
		1 & d0 \\
		0 & 1 
	\end{pmatrix}
	\label{Eq:optic_system}
\end{equation}

where $\mathbb{T}_{d0}$ and $\mathbb{T}_s$ denote the translation matrices for the ray in air before and after the lens. $\mathbb{L}_{f_0}$ is the lens matrix. In order to model the acquisition of the photon $j$ by the pixel $(x_s,y_s)$ of the camera sensor, we applied the system matrix $\mathbb{S}$ on packets of photons exiting the tissue:

\begin{equation}
	\begin{pmatrix}
		x_s \\
		\theta_{x,s}
	\end{pmatrix}  =  \mathbb{S} 
	\begin{pmatrix}
		x \\
		\theta_x
	\end{pmatrix}
	\quad , \quad
	\begin{pmatrix}
		y_s \\
		\theta_{y,s}
	\end{pmatrix}  =  \mathbb{S} 
	\begin{pmatrix}
		y \\
		\theta_y
	\end{pmatrix}
	\label{Eq:ray_tracing}
\end{equation}

The diffuse reflectance is calculated for each pixel, either on the tissue surface or on the camera sensor \cite{Eq_dr2,Eq_dr1}:

\begin{equation}
	\phi(x,y,\lambda) = \frac{\sum_{j=1}^{N_{photons}} w_j(x,y,\lambda) }{N_{photons} . A_p},
	\label{Eq:Diffuse_reflectance}
\end{equation}

where $N_{photons}$ is the number of packets of photons detected at the pixel $(x,y)$, $A_p$ is the area (in $mm^2$) of the pixel and $w_j$ is the weight of the detected packet of photons $j$ defined by:

\begin{equation}
	w_j(x,y,\lambda) = \Pi_{i=1}^{N_{regions}} \exp \left( -\mu_{a,i}(\lambda) . ppl_{i,j}(x,y,\lambda) \right).
\end{equation}

$N_{regions}$ is the number of regions modelled in the tissue (grey matter, large and small blood vessels), $\mu_{a,i}(\lambda)$ (in $mm^{-1}$) is the absorption coefficient for the wavelength $\lambda$ of the region $i$ and $ppl_{i,j}(x,y,\lambda)$ is the partial pathlength (in $mm$) of the detected packet of photon $j$ that travelled in the region $i$ for the wavelength $\lambda$. The mean pathlength of travelled photons in the tissue is calculated with Eq.~\eqref{Eq:Mean_path} \cite{Eq_dr1}:

\begin{equation}
	L(x,y) =  \frac{\sum_{j=1}^{N_{photons}} \sum_{i=1}^{N_{regions}} ppl_{i,j}(x,y,\lambda) .  w_j(x,y,\lambda)}{\sum_{j=1}^{N_{photons}} w_j(x,y,\lambda)}
	\label{Eq:Mean_path}
\end{equation}

Finally, we applied an adaptive non-local means on reconstructed images to improve the signal to noise ratio of the Monte Carlo simulations \cite{MC_denoiser} and we performed a linear interpolation to calculate $\phi$ and $L$ values on the wavelength range $400$ nm to $1000$ nm by steps of $1$ nm.

\subsection{Identification of the optimal wavelength for hemodynamic and metabolic monitoring} \label{sec:optim}

Using the digital instrument simulator (see section \ref{sec:digital_instrument_simulator}), we developed an optimization procedure based on a genetic algorithm (differential evolution) to identify the best wavelength combination in the visible and near infrared range to quantify concentration changes in $HbO_2$, $Hb$, $oxCCO$, $oxCytb$ and $oxCytc$, see Fig.~\ref{fig:schema_optim}.

\subsubsection{Model of cerebral activation}
\begin{figure}[H]
	\centering
	\includegraphics[width=1.0\linewidth]{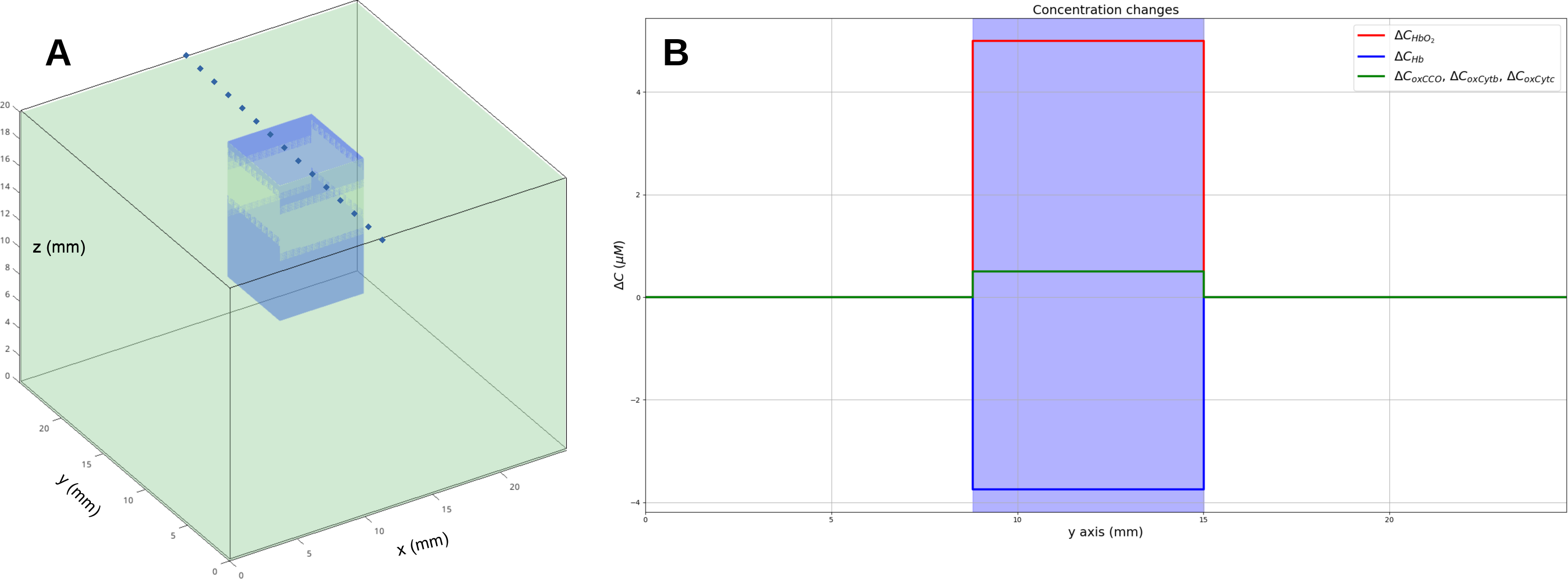}
	\caption{Monte Carlo model used for the optimization procedure. A - Volume of non activated grey matter (green) and activated grey matter (blue). B - Spatial profile of the modelled concentration changes $\Delta C_{GT}$ indicated by the dotted line in figure A.}
	\label{fig:model_optim}
\end{figure}

Data used for the optimization procedure were generated with the digital instrument simulator. A simple model was considered, that includes non activated and activated grey matter, see Fig.~\ref{fig:model_optim} (A). In Fig.~\ref{fig:model_optim} (B), we modelled a cerebral activity in the activated grey matter by changing the concentration of chromophores relative to their nominal values. The modelled concentration changes, that are the ground truth in the optimization procedure are indicated $\Delta C_{GT}$ in the rest of the manuscript. A simple temporal perturbation of $\mu_a$ was performed: rest (nominal values, see table \ref{tab:mua}) and activity (nominal value $+$ $\Delta C$). The $\Delta C$ values were $+5$ $\mu M$ and $-3.75$ $\mu M$ for $HbO_2$ and $Hb$, respectively, and $+0.5$ $\mu M$ for $oxCCO$, $oxCytb$ and $oxCytc$, mirrored by a concentration changes of $-0.5$ $\mu M$ for $redCCO$, $redCytb$ and $redCytc$ (the concentration of cytochromes does not change during the cerebral activity).

\subsubsection{Optimization procedure}

%optimization on Hb + HbO2
%optimization on Hb + HbO2 + oxCCO
%optimization on Hb + HbO2 + oxCCO + oxCytb + oxCyt c

\begin{figure}[H]
	\centering
	\includegraphics[width=1.0\linewidth]{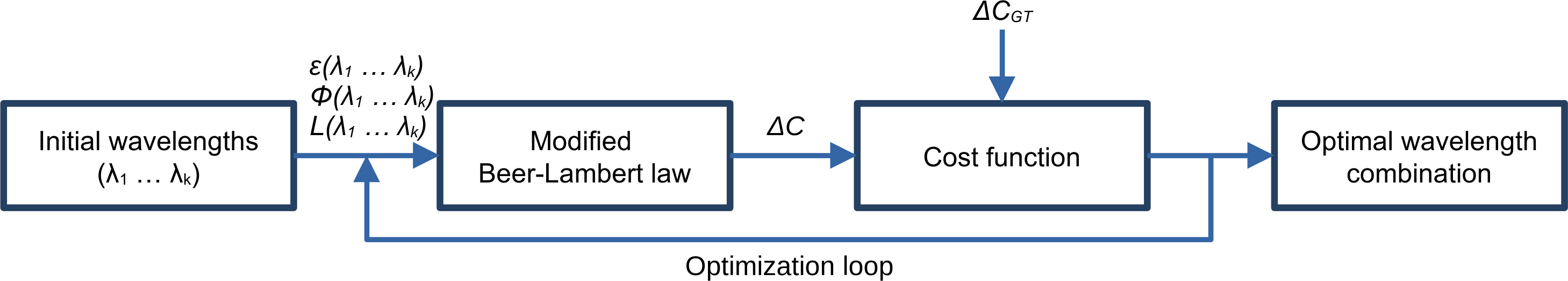}
	\caption{Flowchart of the optimization method, including definitions of the variables used in the procedure.}
	\label{fig:schema_optim}
\end{figure}

The most straightforward method to determine k optimal wavelengths for cerebral monitoring is to test all possible k-element combinations out of $601$ wavelengths (from $400$ nm to $1000$ nm) and to identify the combination that produces the smallest estimation errors when compared against the ground truth $\Delta C_{GT}$. As the investigated spectral range is large, such an exhaustive search results in a large time complexity. Thus, we proposed to use a genetic algorithm \cite{genetic_algo} in order to determine the k optimal wavelengths through an optimization procedure, see Fig.~\ref{fig:schema_optim}. We used the differential evolution method, that optimizes a problem by iteratively trying to improve a candidate solution with regard to the ground truth. We used the function differential$\_$evolution from the Python library Scipy (v1.10.0) \cite{Scipy}.

The optimization procedure consists in finding the best k-element combinations out of $601$ wavelengths that minimize a cost function. This function takes as input a wavelength array, whose values are limited to the simulated spectral range (between $400$ and $1000$ nm). The cost function returns the mean square error $MSE$ computed between noisy concentration changes measured with the modified Beer-Lambert law and the ground truth $\Delta C_{GT}$:

\begin{equation}
	MSE = \sum_{i=1}^{N_{Noise}} \frac{\Delta C_{mes}^{Noise} - \Delta C_{GT}}{N_{Noise}}, 
	\label{Eq:cost}
\end{equation}

where $N_{Noise}=1000$ is the number of iteration for the noise addition. Noise was added to the simulated intensities to identify the k-wavelengths the most robust to the noise. The intensities were converted into concentration changes $\Delta C_{mes}^{Noise}$ measured with the modified Beer-lambert law:

\begin{equation}
	\resizebox{1 \textwidth}{!} {%
		$
		\begin{pmatrix}
			\Delta A(\lambda_1)^{Noise} \\
			\vdots \\
			\Delta A(\lambda_1)^{Noise}
		\end{pmatrix}
		= 
		\begin{pmatrix}
			L(\lambda_1).\epsilon_{HbO_2}(\lambda_1) & L(\lambda_1).\epsilon_{Hb}(\lambda_1) &
			L(\lambda_1).\epsilon_{ox-redCCO}(\lambda_1) &
			L(\lambda_1).\epsilon_{ox-redCytb}(\lambda_1) &
			L(\lambda_1).\epsilon_{ox-redCytc}(\lambda_1) \\
			\vdots & \vdots & \vdots & \vdots & \vdots \\
			L(\lambda_k).\epsilon_{HbO_2}(\lambda_k) & L(\lambda_k).\epsilon_{Hb}(\lambda_k) &
			L(\lambda_k).\epsilon_{ox-redCCO}(\lambda_k) &
			L(\lambda_k).\epsilon_{ox-redCytb}(\lambda_k) &
			L(\lambda_k).\epsilon_{ox-redCytc}(\lambda_k)
		\end{pmatrix}
		\begin{pmatrix}
			\Delta C_{HbO_2}^{Noise} \\
			\Delta C_{Hb}^{Noise} \\
			\Delta C_{oxCCO}^{Noise} \\
			\Delta C_{oxCytb}^{Noise} \\
			\Delta C_{oxCytc}^{Noise}
		\end{pmatrix}$%
	}
	\label{Eq:MBLL}
\end{equation}

In \eqref{Eq:MBLL}, $\epsilon_n(\lambda_1)$ is the molar extinction coefficient of the chromophore $n$ (in $M^{-1}.mm^{-1}$). $\Delta A(\lambda_1)^{Noise}$ is the noisy attenuation change measured for $\lambda_1$, such as:

\begin{equation}
	\Delta A(\lambda_1)^{Noise} = \log_{10} \left( \frac{\phi_{rest}(\lambda_1) + \gamma }{\phi_{activity}(\lambda_1) + \gamma} \right),
	\label{Eq:Delta_A}
\end{equation}

where $\gamma$ is a zero-mean Gaussian noise whose standard deviation was the same for each wavelength: $\sigma = \frac{\phi_{rest}}{SNR}$. With $\phi_{rest}$, the diffuse reflectance simulated during rest and $SNR=400$ the signal to noise ratio of the instrument. This value corresponds to the experimental devices we used in a previous study \cite{numeric_phantom_caredda}, but this could be adapted for another experimental device. The concentration changes were obtained by matrix inversion in the least square sense.

\section{Results}

\subsection{Digital instrument simulator}

\begin{figure}[H]
	\centering
	\includegraphics[width=1.0\linewidth]{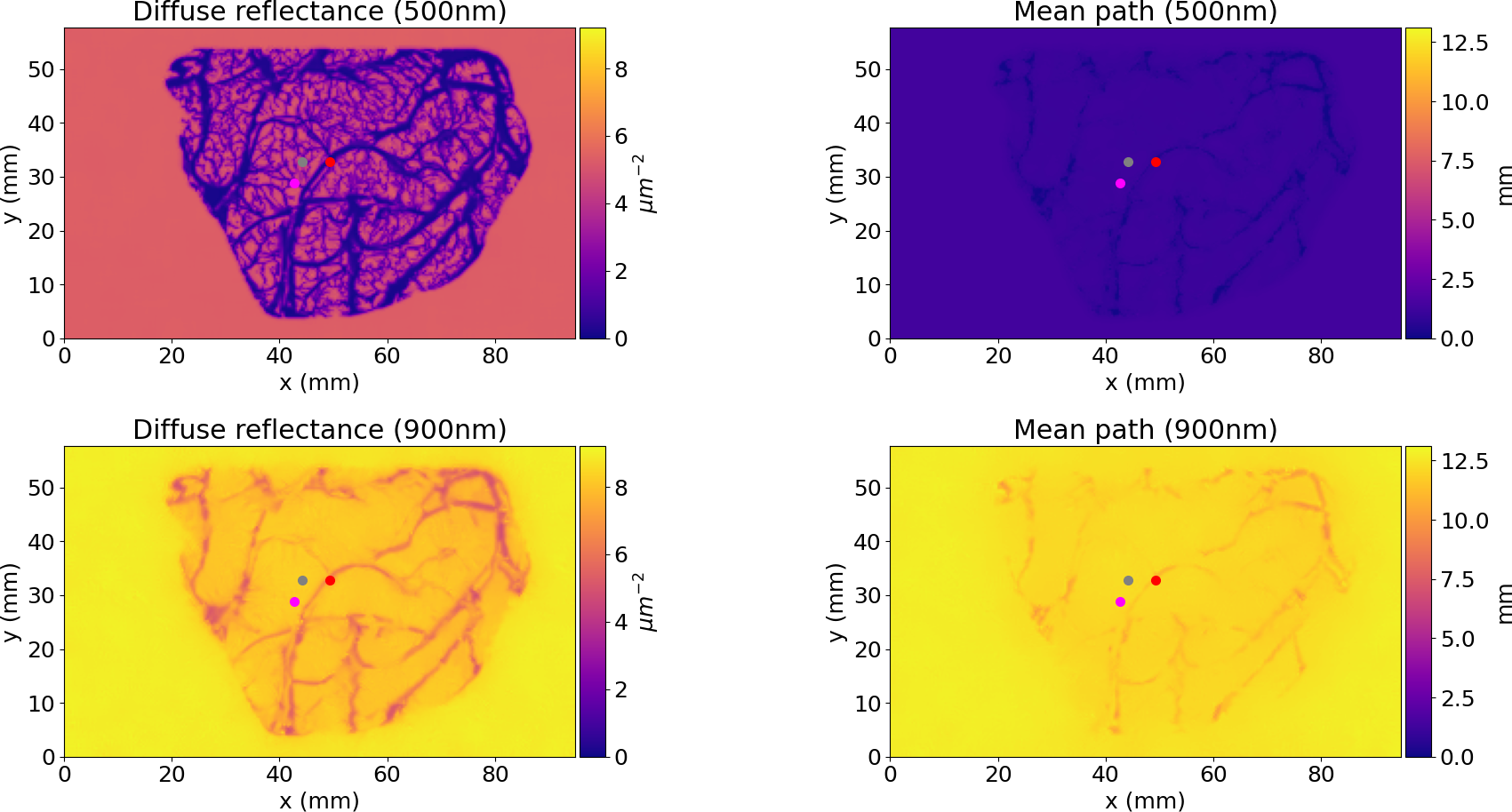}
	\caption{Images of diffuse reflectance and mean path length reconstructed with a $5 \times 5$ binning at $500$ nm and $900$ nm. The large and small blood vessels are considered to be arteries.}
	\label{fig:dr_mp_images}
\end{figure}

In Fig.~\ref{fig:dr_mp_images}, we represented the images of diffuse reflectance and mean path length at $500$ nm and $900$ nm reconstructed at the surface of the tissue considering that large and small blood vessels are arteries. The images were reconstructed with a $5 \times 5$ binning which leads to a resolution of $365$ $\mu m$. The red, grey and magenta points indicate the position of points for a large blood vessel, for grey matter and for a small blood vessel, respectively.

\begin{figure}[H]
	\centering
	\includegraphics[width=1.0\linewidth]{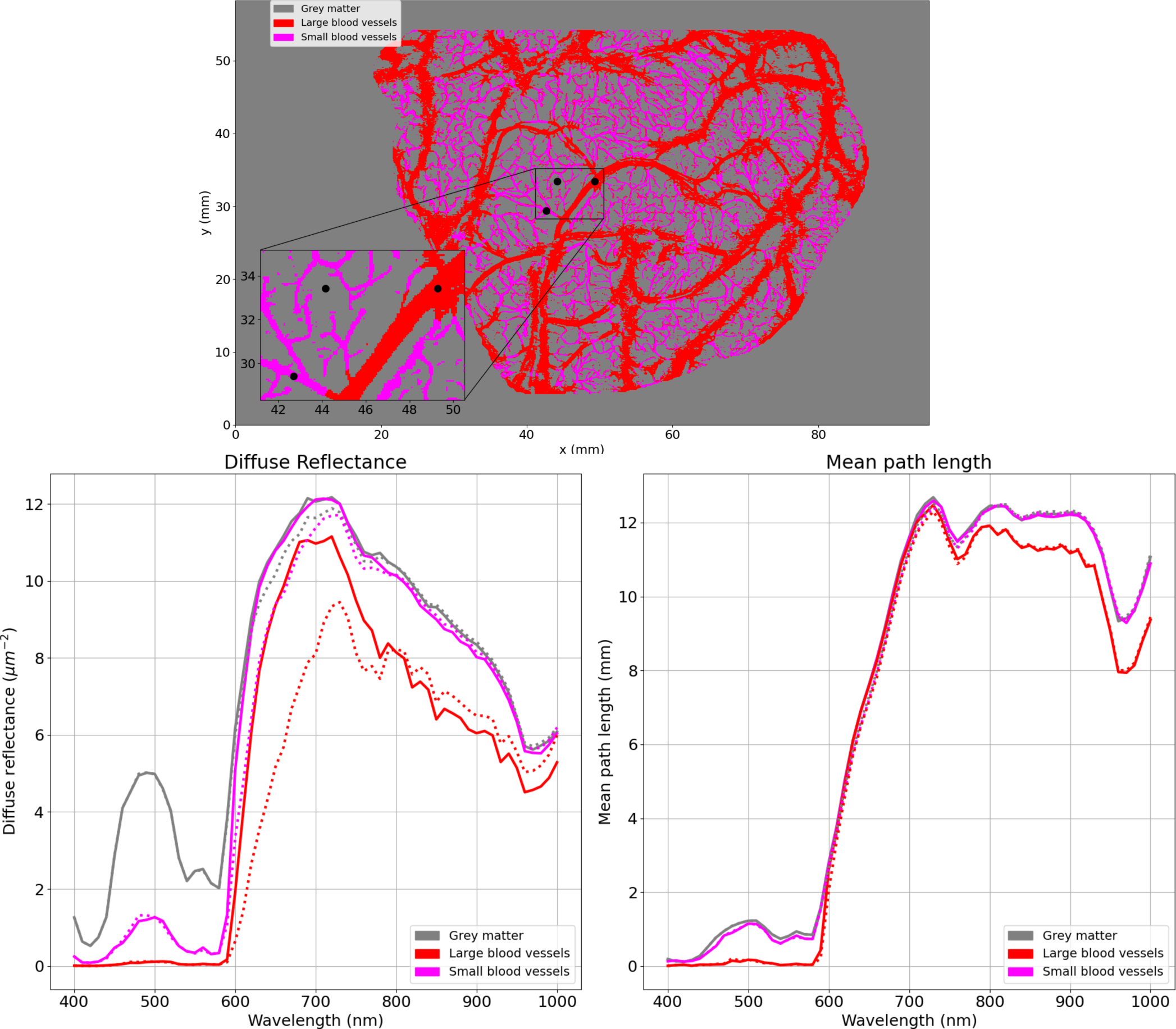}
	\caption{Diffuse reflectance and mean path length spectra measured on grey matter, a large blood vessel and a small blood vessel. Solid and dotted lines indicate that large and small blood vessels are considered arteries and veins, respectively.}
	\label{fig:spectra_digital_simulator}
\end{figure}

In Fig.~\ref{fig:spectra_digital_simulator}, we represented the diffuse reflectance and mean pathlength spectra measured at the level of the points identified in Fig.~\ref{fig:dr_mp_images}. Solid and dotted lines indicate that large and small blood vessels are considered arteries and veins, respectively. For both cases, we can observe the effect of the blood vessels on the simulated quantities. For wavelengths lower than $600$ nm, the diffuse reflectance was almost $0$ $\mu m^{-2}$ on the large blood vessels, this means that hardly any photons left the large blood vessel. Diffuse reflectance values increased when measuring small blood vessels, indicating a greater contribution from outgoing photons. The higher values can be found for grey matter regions. For wavelengths higher than $600$ nm, diffuse reflectance values were almost the same for grey matter and small blood vessels and lower values can be found at the level of the large blood vessels. The mean path length spectra measured for grey matter and the small blood vessels were almost the same for all wavelengths, but lower values were obtained for the large blood vessels. We can observe that the values of the mean pathlength are not impacted if large and small blood vessels are considered arteries or veins. However, we can observe differences between the diffuse reflectance spectra measured at the level of the large blood vessels for wavelengths higher than $600$ nm. There is a drop in intensity in the red and near infrared for veins compared to arteries. This corresponds to the fact that veins appear darker or bluer than arteries.

%In Fig.~\ref{fig:effect_sensor}, we represented the diffuse reflectance images reconstructed on a sensor of $400 \times 320$ pixels of size $6 \times 4.8$ mm. The lens of the optics system had a focal length $f_0 = 30$ mm and its distance to the sensor was $32.43$ mm. This distance was calculated with the Descartes formula by considering that the perfect setting was obtained with a working distance $400$ mm (distance lens to tissue). Two working distances were considered: $400$ mm and $402$ mm. We can see the blurring effect of a non-ideal optics on the image quality when using a working distance of $402$ mm. We can also see the effect of the lens on the field of view, as the image formed by the lens on a real sensor is cropped compared to the entire surface of the brain.
%
%
%\begin{figure}[H]
%	\centering
%	\includegraphics[width=1.0\linewidth]{img/effect_sensor.png}
%	\caption{Diffuse reflectance image at $500$ nm reconstructed on a camera sensor using a working distances of $400$ nm and $402$ mm.}
%	\label{fig:effect_sensor}
%\end{figure}

\subsection{Identification of the optimal wavelength for hemodynamic and metabolic monitoring}

\begin{figure}[H]
	\centering
	\includegraphics[width=1.0\linewidth]{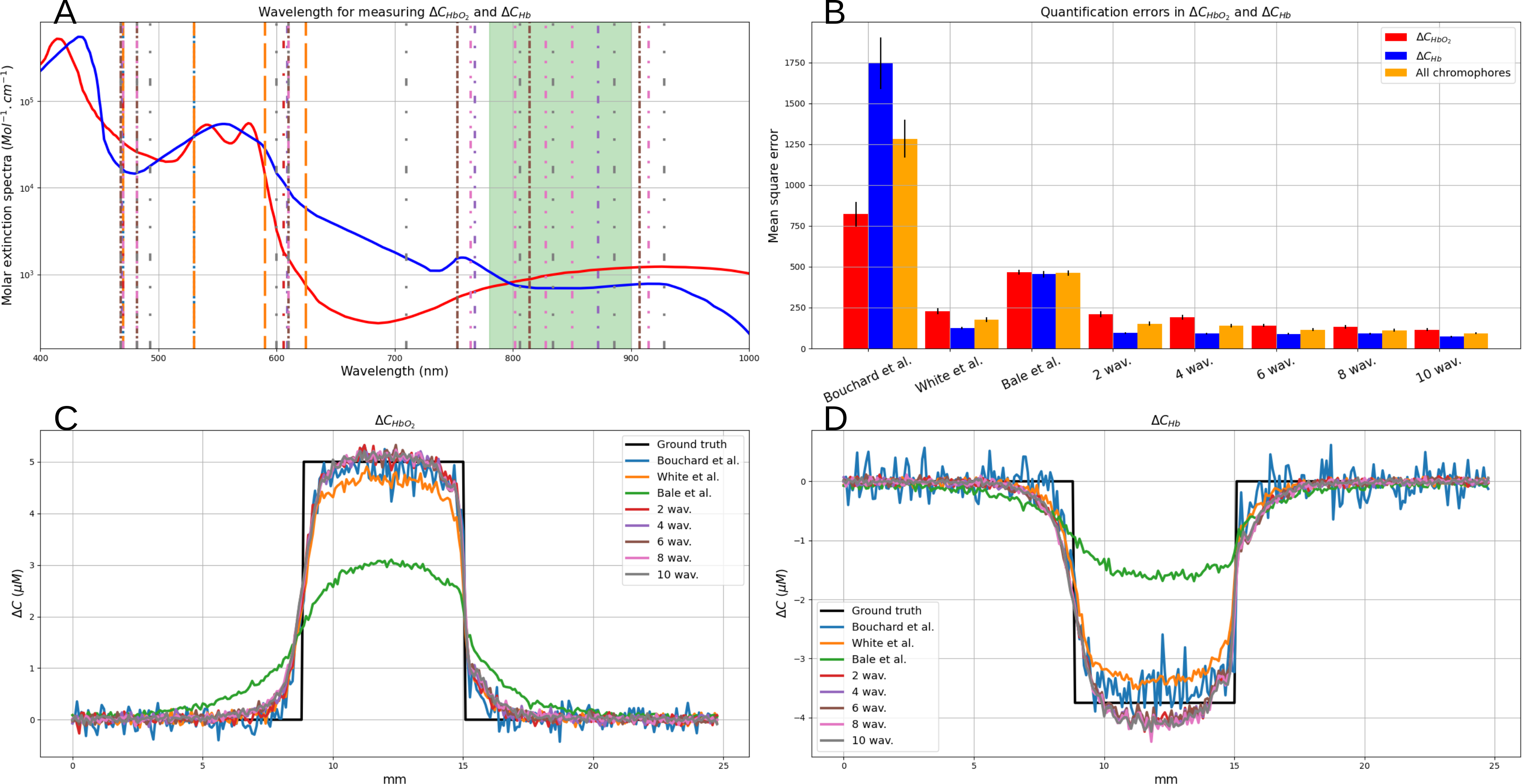}
	\caption{(A) Optimal wavelength for monitoring $C_{HbO_2}$ and $C_{Hb}$ changes in activated grey matter. The quantification errors (B) as well as the concentration changes (C) and (D) are also plotted. Several configuration are represented: $2$ wavelengths proposed by Bouchard et al. \cite{Bouchard}, $4$ wavelengths proposed by White et al. \cite{White}, a broadband spectra proposed by Bale et al.\cite{review_oxCCO,CYRIL} and the combination of $2$, $4$, $6$, $8$ and $10$ wavelengths identified with our optimization procedure.}
	\label{fig:results_optim_hemo}
\end{figure}

%Hemodynamic
In Fig.~\ref{fig:results_optim_hemo}, we represented the optimal combination of $2$, $4$, $6$, $8$ and $10$ wavelengths for monitoring $C_{HbO_2}$ and $C_{Hb}$ changes in activated grey matter. In Fig.~(A), these wavelengths are plotted with vertical lines on the extinction molar spectra of $HbO_2$ and $Hb$ \cite{UCL_spectra}. We also represented the configuration of $2$ wavelengths used by Bouchard et al. \cite{Bouchard}, the $4$ wavelengths used by White et al. \cite{White} and the broadband spectral range between $780$ nm and $900$ nm used by Bale et al. \cite{CYRIL,review_oxCCO} and Arifler et al. \cite{Arifler}. The quantification errors related to these configurations are plotted in Fig.~(B). The bars represent the $MSE$ averaged over 1000 noisy measurements, see Eq.~\eqref{Eq:cost}, and the vertical line the standard deviation in the error. Finally, the concentration changes in $\Delta C_{HbO_2}$ and $\Delta C_{Hb}$ averaged over the $1000$ noisy measurements are plotted in Fig.~(C), and (D), respectively.

Contrary to the configurations proposed by Bouchard et al. and White et al., the wavelengths identified with the optimization procedure do not integer the haemoglobin isobestic point at $530$ nm. The wavelengths were located in the visible and near infrared range. Wavelengths were identified before $500$ nm, where $HbO_2$ absorption predominates, between $590$ nm and $730$ nm where $Hb$ absorption predominates and on either side of the isobestic point at $800$ nm. The minimum errors in $\Delta C_{HbO_2}$ and $\Delta C_{Hb}$ are obtained with the combination of $10$ wavelengths calculated with our optimization procedure. Among the literary configurations, White's spectral configuration minimizes errors in $\Delta C_{HbO_2}$ and $\Delta C_{Hb}$. Errors in $\Delta C_{HbO_2}$ and $\Delta C_{Hb}$ obtained with the optimal group $10$ wavelengths were $50\%$ and $42\%$ lower than those obtained with White's spectral configuration. The mean and standard deviation of the MSE decreases as more spectral bands are considered. For all wavelengths, the mean MSE is lower than that computed with White's spectral configuration. The standard deviation is less than or equal to that obtained with White's configuration from $4$ wavelengths.\\

\begin{figure}[H]
	\centering
	\includegraphics[width=1.0\linewidth]{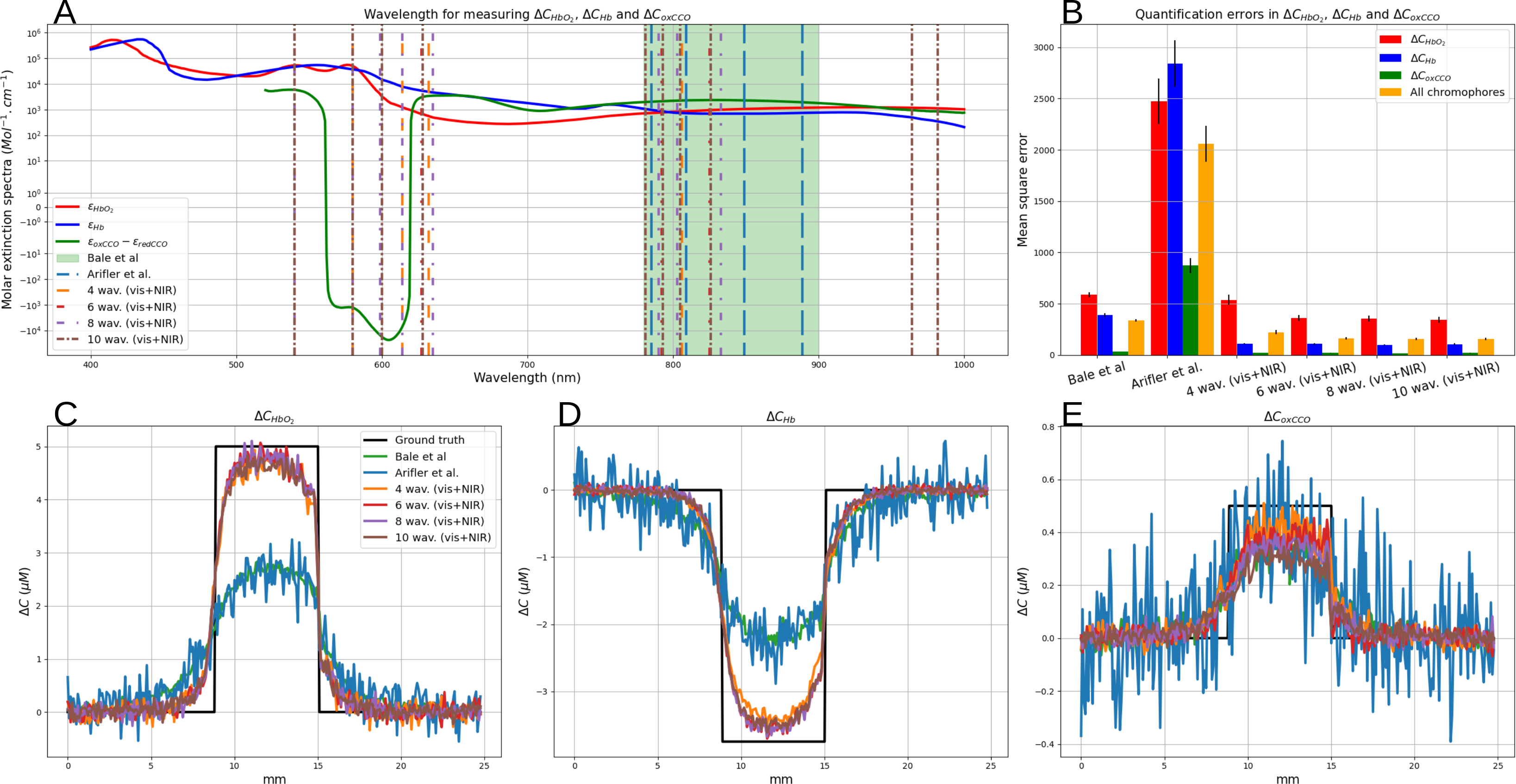}
	\caption{(A) Optimal wavelength for monitoring $C_{HbO_2}$, $C_{Hb}$ and $C_{oxCCO}$ changes in activated grey matter. The quantification errors (B) as well as the concentration changes (C), (D), (E) are also plotted. Several configuration are represented: $4$ wavelengths proposed by Arifler et al. \cite{Arifler}, a broadband spectra proposed by Bale et al.\cite{review_oxCCO,CYRIL} and the combination of $4$, $6$, $8$ and $10$ wavelengths identified with our optimization procedure.}
	\label{fig:results_optim_hemo_oxCCO}
\end{figure}

% Hemo oxCCO
In Fig.~\ref{fig:results_optim_hemo_oxCCO}, we represented the optimal combination of $4$, $6$, $8$ and $10$ wavelengths for monitoring $C_{HbO_2}$, $C_{Hb}$ and $C_{oxCCO}$ changes in activated grey matter. In Fig.~(A), these wavelengths are plotted with vertical lines on the extinction molar spectra of $HbO_2$, $Hb$ and $oxCCO-redCCO$ \cite{UCL_spectra}. We represented the configuration of $4$ wavelengths proposed by Arifler et al. \cite{Arifler} and the broadband spectral range used by Bale et al. \cite{CYRIL,review_oxCCO}. The quantification errors are represented in Fig.~(B) and the measurements of $\Delta C_{HbO_2}$, $\Delta C_{Hb}$ and $\Delta C_{oxCCO}$ are  plotted in Fig.~(C), (D) and (E), respectively.

The minimum errors in $\Delta C_{HbO_2}$, $\Delta C_{Hb}$ and $\Delta C_{oxCCO}$ are obtained with the combination of $10$ wavelengths calculated with our optimization procedure. With this digital instrument simulator, we modelled an exposed cortex. The simulated light is not absorbed by the skin or the skull of the patient, and the visible light can be collected by the camera. Thus, the optimization procedure can take profit of the peaks of oxidized and reduced $CCO$ extinction spectra in the visible and the near infrared range to  monitor the three chromophores. The Errors in $\Delta C_{HbO_2}$, $\Delta C_{Hb}$ and $\Delta C_{oxCCO}$ obtained with this configuration were $40\%$, $72\%$ and $42\%$ lower than those obtained with a broadband spectra between $780$ nm and $900$ nm.  For all chromophores, the standard deviation of the MSE decreases as more spectral bands are considered. For $\Delta C_{HbO_2}$, the mean MSE decreases with the number of wavelengths, while the mean errors for $\Delta C_{Hb}$ and $\Delta C_{oxCCO}$ remain constant. For all wavelengths, the mean MSE is lower than that computed with Bale's spectral configuration. For $\Delta C_{Hb}$ and $\Delta C_{oxCCO}$, the standard deviation is less than or equal to that obtained with Bale's configuration. However, for $\Delta C_{HbO_2}$, the standard deviation is higher than that obtained with Bale's configuration. \\

\begin{figure}[H]
	\centering
	\includegraphics[width=1.0\linewidth]{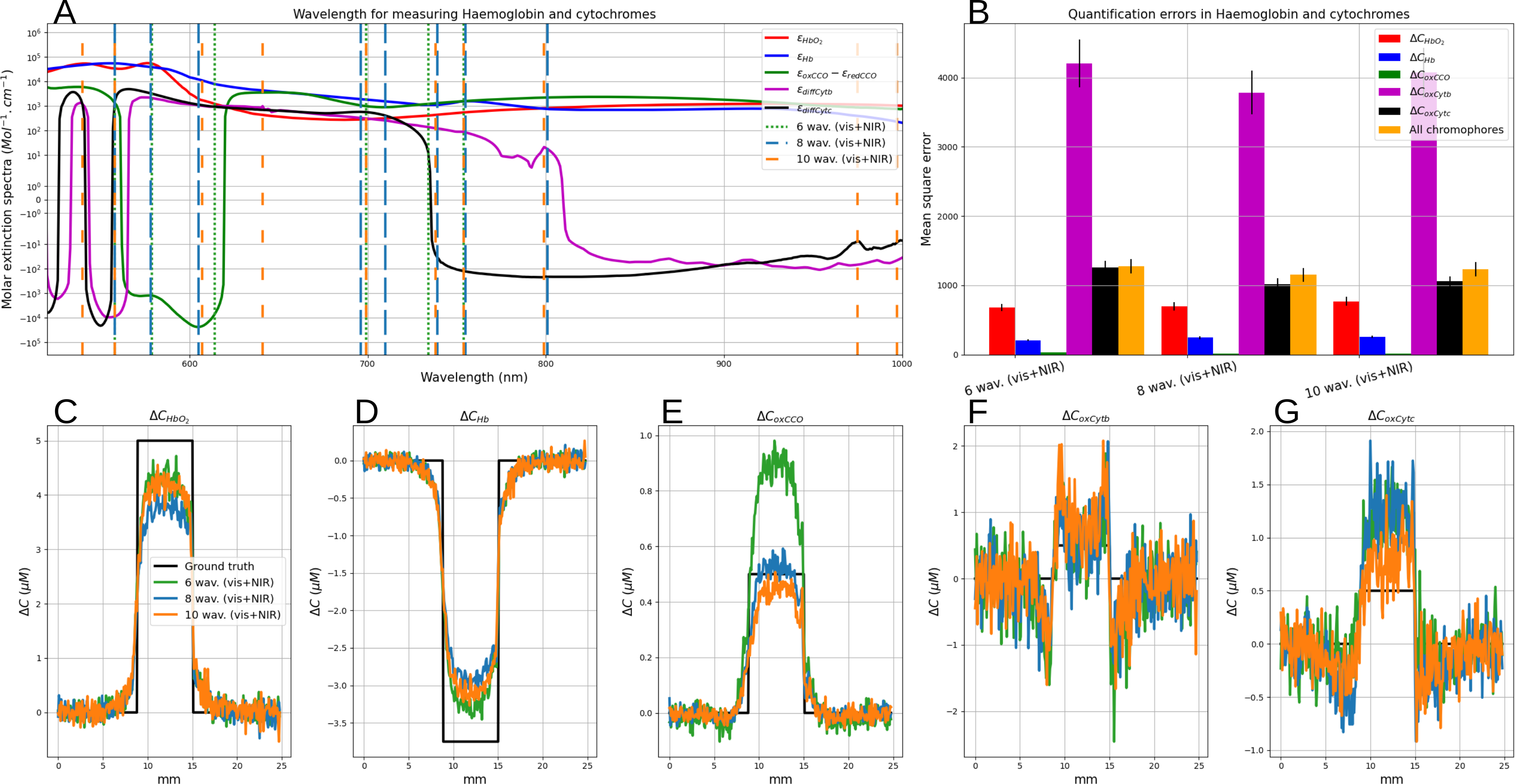}
	\caption{(A) Optimal wavelength for monitoring $C_{HbO_2}$, $C_{Hb}$, $C_{oxCCO}$, $C_{oxCytb}$ ad $C_{oxCytc}$ changes in activated grey matter. The quantification errors (B) as well as the concentration changes (C), (D), (E), (F) and (G) are also plotted.}
	\label{fig:results_optim_hemo_Cyt}
\end{figure}

% All
In Fig.~\ref{fig:results_optim_hemo_Cyt}, we represented the optimal combination of $6$, $8$ and $10$ wavelengths for monitoring $C_{HbO_2}$, $C_{Hb}$, $C_{oxCCO}$, $C_{oxCytb}$ ad $C_{oxCytc}$ changes in activated grey matter. These wavelengths are plotted with vertical lines on the extinction molar spectra in Fig.~(A). The quantification errors are represented in Fig.~(B) and the concentration changes are plotted in Fig.~(C), (D), (E), (F) and (G).

The minimum errors to resolve all chromophores are obtained with the combination of $8$ and $10$ wavelengths. The Errors in $\Delta C_{HbO_2}$ and $\Delta C_{Hb}$ were higher than those obtained in Fig.~\ref{fig:results_optim_hemo_oxCCO}. However, with the consideration of cytochromes b and c in the modified Beer-Lambert law, the errors in $\Delta C_{oxCCO}$ were $41 \%$ lower than those obtained in Fig.~\ref{fig:results_optim_hemo_oxCCO}. The noise in the measurements of $\Delta C_{oxCytb}$ and $\Delta C_{oxCytc}$ is very important, making it difficult to differentiate between activated and non-activated grey matter.\\

%%%%%%%%%%%%%%%

\begin{figure}[H]
	\centering
	\includegraphics[width=1.0\linewidth]{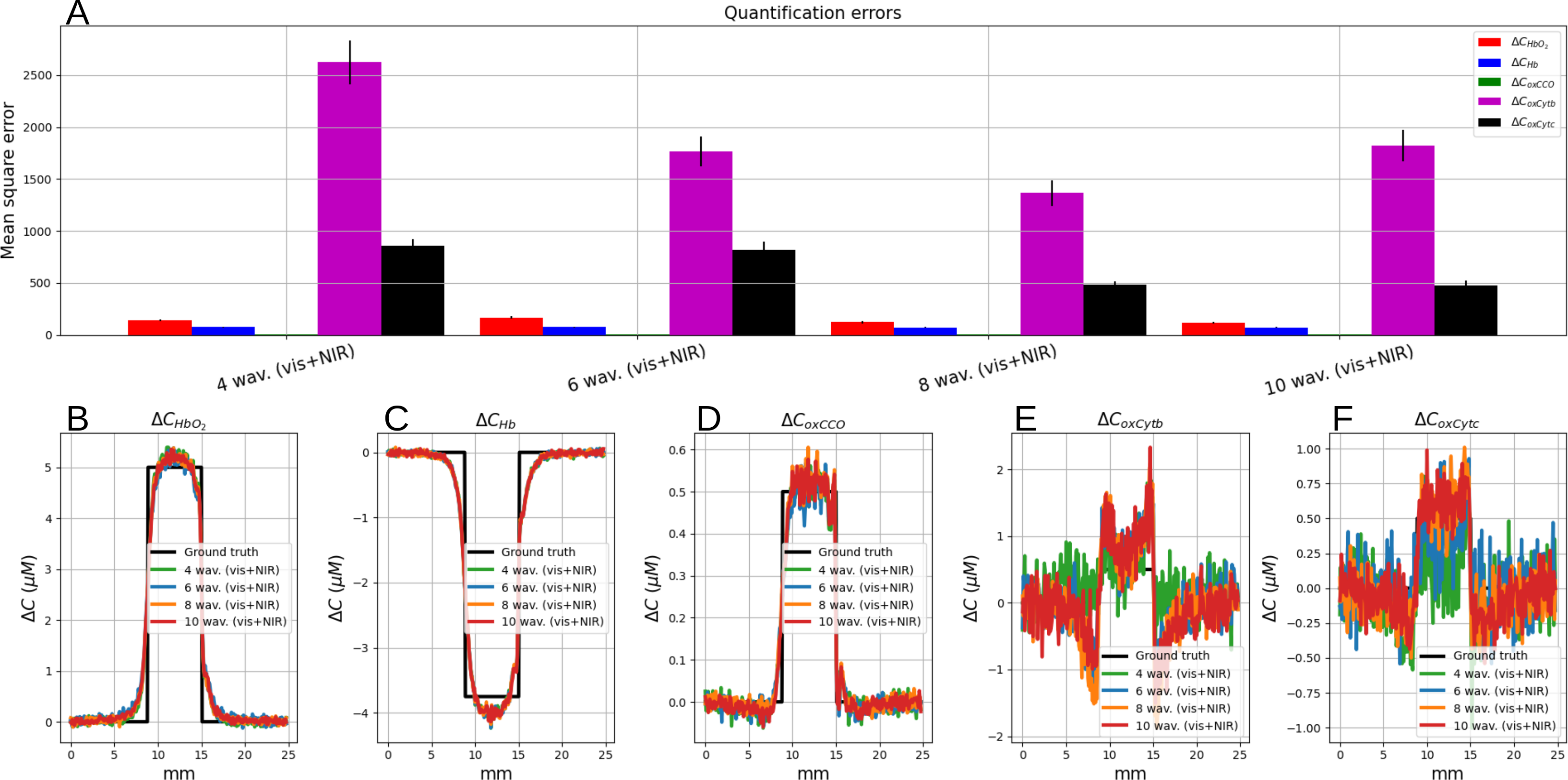}
	\caption{Quantification errors (A) and concentration changes obtained with the optimal combination of $4$, $6$, $8$ and $10$ wavelengths for a separate quantification. The concentration changes in $C_{HbO_2}$, $C_{Hb}$, $C_{oxCCO}$, $C_{oxCytb}$ and $C_{oxCytc}$ are plotted in Fig. (B), (C), (D), (E) and (F), respectively.}
	\label{fig:results_optim_separate1}
\end{figure}

\begin{figure}[H]
	\centering
	\includegraphics[width=1.0\linewidth]{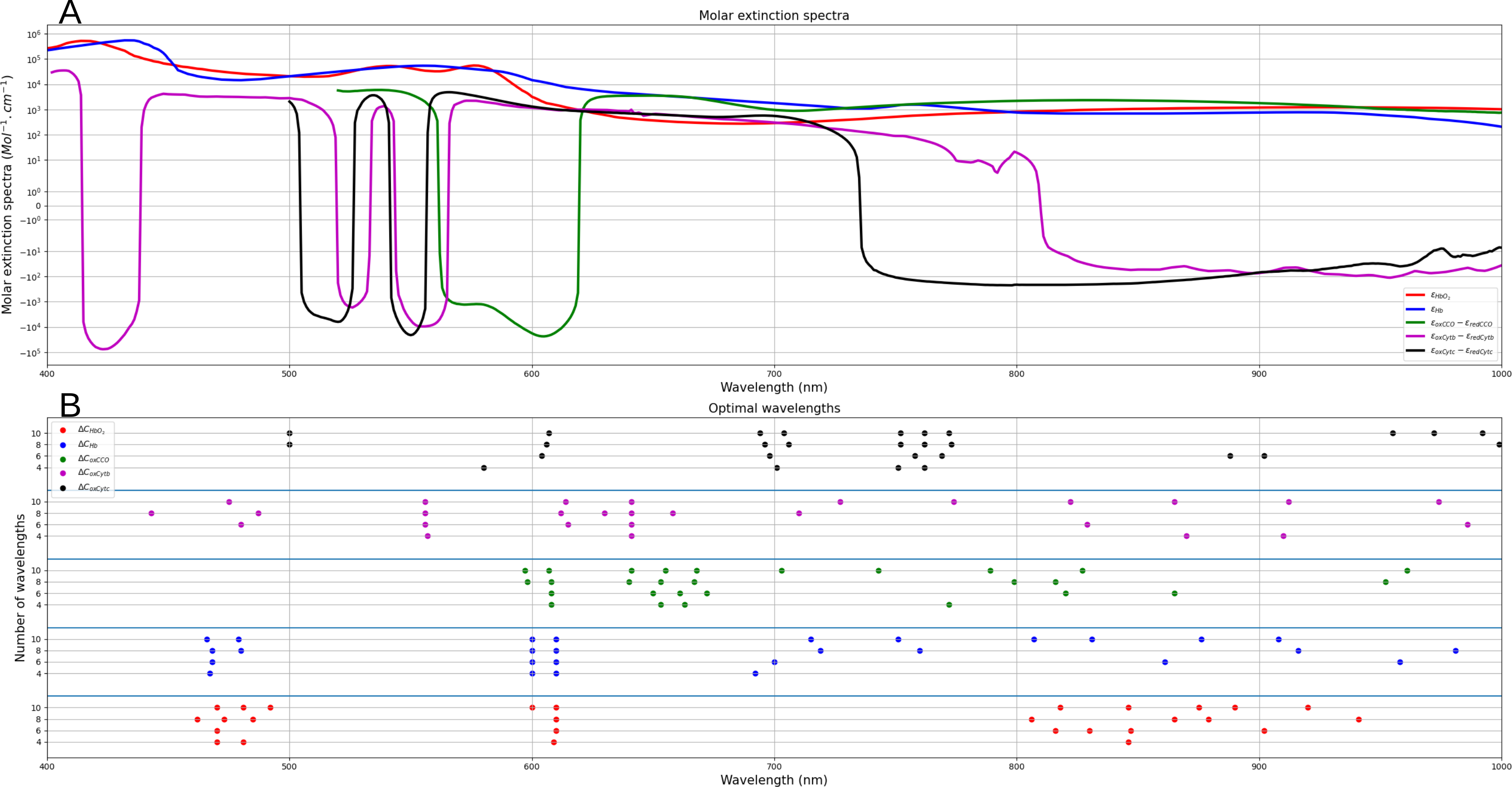}
	\caption{Molar extinction coefficient (A) and optimal wavelength for monitoring $C_{HbO_2}$, $C_{Hb}$, $C_{oxCCO}$, $C_{oxCytb}$ and $C_{oxCytc}$ changes separately (B). The optimal combination of $4$, $6$, $8$ and $10$ wavelengths are represented by colored dots.}
	\label{fig:results_optim_separate2}
\end{figure}

As proposed in our previous study \cite{numeric_phantom_caredda}, we can use different wavelength combinations for separately resolving haemoglobin and cytochrome changes see Fig.~\ref{fig:results_optim_separate1} and Fig.~\ref{fig:results_optim_separate2}. For instance, if we want to identify the optimal wavelength to resolve $C_{oxCytb}$ changes only, the optimization procedure can be run to minimize $\Delta C_{oxCytb}$ values by only considering three chromophores in the modified Beer-Lambert law ($HbO_2$, $Hb$ and $oxCytb$, see Eq.~\eqref{Eq:MBLL}).

With this approach, we represented in Fig.~\ref{fig:results_optim_separate1} (A) the quantification errors obtained with the optimal wavelength combination to separately resolve $HbO_2$, $Hb$, $oxCCO$, $oxCytb$ and $oxCytc$ changes. We also plotted the concentration changes in $HbO_2$, $Hb$, $oxCCO$, $oxCytb$ and $oxCytc$ in Fig. (B), (C), (D), (E) and (F).

In Fig.~\ref{fig:results_optim_separate2} (B), we represented the optimal wavelength combination of $4$, $6$, $8$ and $10$ wavelengths for a separate quantification. The separate optimization procedure leads to different spectral configuration for each chromophore.

For all chromophores (expect $oxCytb$), the mean MSE decreases with the number of wavelengths. We obtained smaller quantification errors when we applied the optimization procedure on each chromophore separately than when we tried to resolve all the chromophores at the same time, see table \ref{tab:Quantification_errors}.

\begin{table}[H]
	\centering
	\resizebox{1 \textwidth}{!} {\begin{tabular}{c|c|c|c|c|c}
			Modified Beer-Lambert law system & $HbO_2$ & $Hb$ & $oxCCO$ & $oxCytb$ & $oxCytc$ \\
			\hline
			Broadband spectra ($780$ - $900$ nm) \cite{CYRIL,review_oxCCO} & $583$ & $395$ & $34$ & & \\
			\hline
			White's configuration \cite{White} & $227$ & $124$ & & & \\
			\hline
			$HbO_2$, $Hb$ and $oxCCO$ ($6$ wavelengths), see Fig.~\ref{fig:results_optim_hemo_oxCCO} & $355$ & $98$ & $17$ & & \\
			\hline
			All chromophores at the same time ($6$ wavelengths), see Fig.~\ref{fig:results_optim_hemo_Cyt}& $743$ &  $247$ &   $13$ & $3969$ & $1055$ \\
			\hline
			All chromophores separately ($6\times 5$ wavelengths), see Fig.~\ref{fig:results_optim_separate2} & $138$ & $72$ & $6$ & $1779$& $818$ \\
			
	\end{tabular}}
	\caption{Average of the quantification errors (MSE) obtained with two Beer-Lambert systems proposed in the literature and optimal combination of $6$ wavelengths.}
	\label{tab:Quantification_errors}
\end{table}

\section{Discussion}

We presented a digital instrument simulator to optimize the development of hyperspectral systems for intraoperative brain mapping studies. The code is publicly available on \href{https://github.com/CCaredda/White-Monte-Carlo/}{GitHub}. This simulator is based on a realistic digital phantom of an exposed cortex computed with white Monte Carlo simulations. We developed a C++ software to reconstruct temporal perturbations of the absorption coefficient using only one simulation. Using this Monte Carlo framework, we also proposed an optimization procedure based on the genetic algorithm to identify the best wavelength combinations in the visible and near infrared range to quantify changes in $HbO_2$, $Hb$, $oxCCO$, $oxCytb$ and $oxCytc$.

%Digital instrument simulator
\subsection{Digital instrument simulator}

The digital instrument allows the modelling of intensity maps collected by a camera sensor as well as the estimation of the mean path length of travelled photons through the tissue. This framework could be used to improve clinical and preclinical optical devices for brain mapping applications. In these studies, the modified Beer-Lambert law is used to monitor chromophore changes in animal \cite{Bouchard,White} or human \cite{caredda,caredda_neuroimage,caredda_resting_state,Pichette} brain. However, measurements are highly subject to quantification errors if incorrect pathlengths are used to resolve chromophore changes. This could be the case in a lot of applications where the inhomogeneities of the optical properties are not taken into consideration. Indeed, a single pathlength is considered for the whole cerebral cortex which is usually estimated with a homogeneous volume of grey matter with Monte Carlo simulations \cite{Bouchard} or using the analytical solution to the diffusion approximation of the radiative transfer equation in a semi-inﬁnite geometry \cite{White,Pichette}. In our study, we proposed to take into account the inhomogeneities of the optical properties with a pixel-wise estimation of the mean pathlength. As we can see on Figs.~\ref{fig:dr_mp_images} and \ref{fig:spectra_digital_simulator}, the blood vessels have a great impact on the pathlength. For large blood vessels having a $2$ mm diameter, the pathlength is almost null in the visible spectra (below $600$ nm) and is lower than the pathlength estimated for grey matter in the red and near infrared range. However, the pathlength measured at the level of the small blood vessels ($\approx 0.2$ mm) is almost the same than that measured at the level of the grey matter. This means that the detected packets of photon are mostly propagated in the grey matter under these small blood vessels. This result is consistent with the results presented by Giannoni et al. \cite{3D_Heterogeneous_Phantom} and indicates that large blood vessels have a large impact on the pathlength.

%Improvement
In this study, we modelled a perfect optical device. The real spectral characteristics of the light source and the sensor could be integrated into the digital simulator by normalizing the estimated quantities with spectral sensitivity curves of the light sources and the camera, as proposed in our previous study \cite{numeric_phantom_caredda}. We modelled a real sensor in term of resolution and definition, but this could be improved with the conversion of the diffuse reflectance into sensor intensities by considering parameters of the optics systems \cite{camera_simu} (lens magnification, lens transmission, quantum efficiency). We can also integrate the effect of the lenses on the image creation such as geometric distortion, vignetting or chromatic aberrations. A lens design model can be used for this purpose \cite{lens_model}. Using this digital instrument simulator we modelled a simple cerebral activation by changing the concentrations of the chromophores in a portion of activated grey matter (see Fig.~\ref{fig:model_optim}). However, we do not model the complex physiologic events related to the neuro-vascular coupling. To answer this limitation, we plan to model the spatio-temporal concentration changes in arteriole, capillary and venous compartments related to blood volume, flow velocity, and oxygen consumption with the implementation of the dynamic model proposed by S. Fantini \cite{dynamic_model_Fantini}.

%Limitations
Although the strength of this digital instrument is to use the white Monte Carlo approach, this method has also a limitation which is the size of the generated files. Indeed, for one wavelength, $5$ GB of data is generated, which lead to $305$ GB for the $61$ wavelengths modelled in our study. To reduce the memory size, the absorption coefficients need to be fixed in the modelled tissue in order to directly estimate the radiative quantities. But with this approach, we cannot reconstruct the temporal changes of the absorption coefficient, so a simulation has to be performed per wavelength and per temporal index, which increases drastically the computation time.

\subsection{Identification of the optimal wavelength for hemodynamic and metabolic monitoring}

Using the data obtained with the digital instrument simulator, we proposed an optimization procedure based on the genetic algorithm to identify the best wavelength combinations in the visible and near infrared range to quantify changes in $HbO_2$, $Hb$, $oxCCO$, $oxCytb$ and $oxCytc$. We added noise to the simulated quantities to identify the wavelength that are the most robust to noise. 

%Hemodynamic
For hemodynamic monitoring, see Fig.~\ref{fig:results_optim_hemo}, we can see that the optimal groups of $2$, $4$, $6$, $8$ or $10$ wavelengths aim to reduce the quantification errors in $\Delta C_{HbO_2}$ and $\Delta C_{Hb}$ of $50 \%$ and $42 \%$ compared to that obtained with the configurations proposed in the literature. Contrary to the spectral configurations proposed by Bouchard et al. \cite{Bouchard} and White et al. \cite{White}, the wavelengths identified with the optimization procedure do not include the haemoglobin isobestic point at $530$ nm. The wavelengths were located in the visible and near infrared range, where $HbO_2$ or $Hb$ absorption predominates. This could be explained by the addition of the noise in the optimization procedure, which makes difficult to interpret the attenuation changes measured at the isobestic points. Moreover, the optimization procedure identify the best wavelengths but do not take into account the bandwidth of the light source spectra in hyperspectral devices based on spectral-scanning technology \cite{review_HSI2}. Taking into account a spectral bandwidth could modify the value of the central wavelengths by a few nm, but would not introduce changes greater than the spectral bandwidth.

%Hemodynamic and metabolic monitoring
The optimization procedure took profit of the peaks of oxidized and reduced cytochrome extinction spectra in the visible and the near infrared range to  monitor hemodynamic and metabolic changes. For hemodynamic and $oxCCO$ monitoring, the optimal wavelengths aim to reduce the quantification errors compared to that obtained with a broadband spectra between with $780$ nm and $900$ nm. In this study, we also proposed to quantify the changes of the cytochrome b and c. We can see in Figs.~\ref{fig:results_optim_hemo_Cyt} that the addition of the cytochromes b and c in the modified Beer-Lambert system help to better resolve $oxCCO$ changes compared to the model based on three chromophores ($HbO_2$, $Hb$ and $oxCCO$), see Fig.~\ref{fig:results_optim_hemo_oxCCO}. Indeed, we can see that the quantification errors in $oxCCO$ are lower when all cytochromes are considered, see table \ref{tab:Quantification_errors}. This result is interesting, as it may help to obtain robust devices to monitor brain metabolism, which could be a hyperspectral device for intraoperative brain mapping or even a NIRS device for bedside monitoring. The monitoring of oxidation state of $CCO$ could help to obtain a more direct biomarker of neuronal activity, to detect brain injuries \cite{oxCCO_brain_injury,review_oxCCO2}, and to better understand the neurovascular coupling \cite{neurovascular_coupling}.

The optimization procedure does not help to resolve changes in cytochrome b and c in a significant way. Indeed, we can see in Figs.~\ref{fig:results_optim_hemo_Cyt} and \ref{fig:results_optim_separate1} that $oxCytb$ and $oxCytc$ measurements are really noisy. A two-samples T-test tells us that changes measured at the level of the activated grey matter are not significantly different from those measured on non activated grey matter ($p_{values}$ are mainly higher than $0.1$). This is maybe due to the noise addition, coupled with the fact that the attenuation changes due to the oxidized cytochrome b and c are rather low compared to those of hemoglobin and $oxCCO$.

%Separate quantification
We also showed that a separate quantification of hemodynamic and $oxCCO$ leads to a better estimation of the chromophore changes, see table \ref{tab:Quantification_errors}. The optimization procedure helps to identify the best wavelength combination in the range $400-1000$ nm that reduce the quantification errors in $HbO_2$, $Hb$, $oxCCO$ of $61 \%$, $29\%$ and $82\%$ compared to the gold standard of 121 wavelengths between $780$ and $900$ nm. The configuration of $6$ wavelengths is: $468$, $482$, $610$, $753$, $814$ and $907$ nm to monitor $HbO_2$ and $Hb$ changes, and $608$, $650$, $661$, $672$, $820$, $865$ nm to monitor $oxCCO$ changes. The separate quantification method should be taken with caution because it distorts the link between chromophores. Moreover, this could lead to incorrect interpretations between the chromophore changes and the physiological status of the patient. However, this method could be interesting for some clinical applications. For example, the method can be used to identify a spectral configuration for a precise monitoring of $Hb$ changes with the idea to define a BOLD-like contrast (Blood Oxygen Level Dependant \cite{neurovascular_coupling}). In intraoperative brain mapping studies, this approach has been proposed by several research group. Authors used a single illumination at $605$ \cite{foptics_605} nm or $610$ nm \cite{foptics_610} combined with a monochrome camera to assess $Hb$ absorption. For these two wavelengths, the authors considered that the signal changes was mainly due to $Hb$ because the absorption related with $HbO_2$ is negligible compared with that of $Hb$ (the ratio between $Hb$ molar extinction coefficient to that of $HbO_2$ is $\approx 6.26$). Although the contribution of HbO2 is small compared with that of Hb, the signal measured with this type of device cannot accurately measure variations in Hb concentration. Our method solves this problem.

\section{Conclusion}

In this paper, we proposed two main contributions. Firstly, we presented a digital instrument simulator to optimize the development of hyperspectral systems for intraoperative brain mapping studies. This simulator is based on a realistic digital phantom of an exposed cortex computed with white Monte Carlo simulations. A C++ software was developed to reconstruct temporal perturbations of the absorption coefficient using only one simulation. Secondly, we presented an optimization procedure based on the genetic algorithm to identify the best wavelength combinations in the visible and near infrared range to quantify changes in $HbO_2$, $Hb$, $oxCCO$, $oxCytb$ and $oxCytc$. With the digital instrument simulator, we can improve the accuracy of actual preclinical and clinical optical devices used in brain mapping applications with the consideration of the impact of the large blood vessels on the pathlength. We also proposed several spectral configurations to monitor hemodynamic and metabolic changes in grey matter that aimed to reduce the quantification error of changes in $HbO_2$, $Hb$, $oxCCO$ compared to the configuration proposed in the literature. This digital instrument simulator and this optimization framework could be used to optimize the design of hyperspectral imaging devices \cite{hyperprobe1,hyperprobe2}, with the objective of transforming current practice towards an all-optical, real-time, quantitative and accurate imaging approach, that could significantly help neurosurgeons, enhance the efficacy of the treatment, and ultimately improve the quality of life and functional outcomes of the patients after the brain tumor resection.

% \disclosures 
\subsection*{Disclosures}
The authors declare no conflicts of interest.

\subsection* {Code, Data, and Materials Availability} 
In support of open science, the code presented in this article is publicly available on \href{https://github.com/CCaredda/White-Monte-Carlo/}{GitHub}.

\subsection* {Acknowledgments}
These works were funded by the European Union’s Horizon Europe research and innovation programme under grant agreement No 101071040 – project HyperProbe; LABEX PRIMES (ANR-11-LABX-0063) of Université de Lyon, within the program “Investissements d’Avenir” (ANR-11-IDEX-0007), operated by the French National Research Agency (ANR); Infrastructures d’Avenir en Biologie Santé (ANR-11-INBS-000), within the program “Investissements d’Avenir” operated by the French National Research Agency (ANR) and France Life Imaging (ANR-11-INBS-0006). FL and IT are supported by UCL, which, as UK participant in Horizon Europe Project HyperProbe is supported by UKRI grant number 10048387. We want to acknowledge the PILoT facility for the support provided for the image acquisition. This material is based upon work done on the ISO 9001:2015 PILoT facility.

%%%%% References %%%%%

\bibliography{report}   % bibliography data in report.bib
\bibliographystyle{spiejour}   % makes bibtex use spiejour.bst

%%%%% Biographies of authors %%%%%

%\vspace{2ex}\noindent\textbf{First Author} is an assistant professor at the University of Optical Engineering. He received his BS and MS degrees in physics from the University of Optics in 1985 and 1987, respectively, and his PhD degree in optics from the Institute of Technology in 1991.  He is the author of more than 50 journal papers and has written three book chapters. His current research interests include optical interconnects, holography, and optoelectronic systems. He is a member of SPIE.

\vspace{1ex}
\noindent Biographies and photographs of the other authors are not available.

\listoffigures
\listoftables

\end{spacing}
\end{document}